\documentclass[aip,jmp,graphicx,reprint]{revtex4-1}
            
\usepackage{graphicx}
\usepackage{amsmath}
\usepackage{amssymb}
\usepackage{mathrsfs}      

\usepackage{hyperref}
\hypersetup{pdftitle={Cohomology of Line Bundles: A Computational Algorithm}, 
      pdfauthor={Ralph Blumenhagen, Benjamin Jurke, Thorsten Rahn, Helmut Roschy}, 
      pdfsubject={Algebraic Geometry, Mathematical Methods in Physics},
      pdfstartview={FitH}, pdfpagelayout={TwoColumnRight},
      bookmarksopen, bookmarksnumbered, bookmarksopenlevel=2}

\newcommand{\beq}{\begin{equation}}  \newcommand{\eeq}{\end{equation}}
\newcommand{\bal}{\begin{aligned}}   \newcommand{\eal}{\end{aligned}}

\newcommand{\CPP}{C\nolinebreak\hspace{-.05em}\raisebox{.4ex}{\tiny\bf +}\nolinebreak\hspace{-.10em}\raisebox{.4ex}{\tiny\bf +}}

\def\IR{\mathbb{R}}
\def\IC{\mathbb{C}}
\def\IP{\mathbb{P}}

\def\cO{\mathcal{O}}
\def\cS{\mathcal{S}}
\def\cP{\mathcal{P}}
\def\cQ{\mathcal{Q}}
\def\cH{\mathcal{H}}
\def\cM{\mathcal{M}}

\def\fh{\mathfrak{h}}
\def\fC{\mathfrak{C}}
\def\fU{\mathfrak{U}}
\def\fV{\mathfrak{V}}

\def\shF{\mathscr{F}}
\def\shG{\mathscr{G}}

\def\shI{\mathscr{I}}

\def\labs{\left\|}
\def\rabs{\right\|}

\DeclareMathOperator{\indlim}{\underrightarrow{\lim}}

\def\clap#1{\hbox to 0pt{\hss#1\hss}}

\def\mrlap{\mathpalette\mathrlapinternal}
\def\mclap{\mathpalette\mathclapinternal}

\def\mathrlapinternal#1#2{%
\rlap{$\mathsurround=0pt#1{#2}$}}
\def\mathclapinternal#1#2{%
\clap{$\mathsurround=0pt#1{#2}$}}

\def\ce{\mathrel{\mathop:}=}  

\def\fto{\longrightarrow}
\def\injto{\lhook\joinrel\relbar\!\!\:\joinrel\rightarrow}
\def\surjto{\relbar\joinrel\twoheadrightarrow}

\def\uspc{${}^\big.$}   

\begin{document}







\title{Cohomology of Line Bundles: A Computational Algorithm} 



\eprint{arXiv:1003.5217}
\preprint{arXiv:1003.5217,
  MPP-2010-32,
	NSF-KITP-10-031}

\author{Ralph Blumenhagen}
\email[]{blumenha@mppmu.mpg.de}
\affiliation{Max-Planck-Institut f\"ur Physik, F\"ohringer Ring 6, 80805 M\"unchen, Germany}
\affiliation{Kavli Institute for Theoretical Physics, Kohn Hall, UCSB, Santa Barbara, CA 93106, USA}

\author{Benjamin Jurke}
\email[]{bjurke@mppmu.mpg.de}
\affiliation{Max-Planck-Institut f\"ur Physik, F\"ohringer Ring 6, 80805 M\"unchen, Germany}
\affiliation{Kavli Institute for Theoretical Physics, Kohn Hall, UCSB, Santa Barbara, CA 93106, USA}

\author{Thorsten Rahn}
\email[]{rahn@mppmu.mpg.de}
\affiliation{Max-Planck-Institut f\"ur Physik, F\"ohringer Ring 6, 80805 M\"unchen, Germany}

\author{Helmut Roschy}
\email[]{roschy@mppmu.mpg.de}
\affiliation{Max-Planck-Institut f\"ur Physik, F\"ohringer Ring 6, 80805 M\"unchen, Germany}


\date{\today}


\begin{abstract}

We present an algorithm for computing line bundle valued cohomology classes over toric varieties. This is the basic starting point for computing massless modes in both heterotic and Type IIB/F-theory compactifications, where the manifolds of interest are complete intersections of hypersurfaces in toric varieties supporting  additional  vector bundles.

\end{abstract}


\maketitle




\tableofcontents

\section{Introduction}

It is clear that the computation of line bundle cohomology classes over $D$-dimensio\-nal varieties is an interesting mathematical question of its own. Our motivation for approaching this problem, however, originates from the appearance of this problem in various classes of compactifications of string theory.

First of all, in supersymmetric compactifications of the heterotic string from ten to four space-time dimensions, the massless spectrum is given by certain cohomology classes $H^i(Y;V)$ of the vector bundle $V$ over a Calabi-Yau three-fold $Y$. The largest class of such Calabi-Yau threefolds is given by complete intersections over toric ambient varieties $X$. For the vector bundles various constructions have been discussed in the literature. The three most prominent ones are, on the one hand, monad and extension constructions, for which the vector bundle is defined via (exact) sequences of direct sums of line bundles. On the other hand, for elliptically fibered Calabi-Yau manifolds, the spectral cover construction provides a large class of stable holomorphic vector bundles. However, in all three cases, the computation of the massless spectrum, i.e.~of vector bundle cohomology classes, boils down to determine the cohomology $H^i(X;L)$ of line bundles $L$ over toric ambient varieties $X$. Therefore, it is pretty obvious that it would be important to have a tool to determine these line bundle cohomology classes in a straightforward way, i.e.~without invoking any manifold dependent tricks.

The same problem also  appears  in other kinds of string compactification like Type IIB orientifolds with B-type D-branes or in F-theory compactifications on Calabi-Yau fourfolds. Here one is confronted for instance with computing line bundle cohomology classes over curves, which arise due to the intersection of seven-brane wrapped divisors on a three-dimensional compact internal manifold. Such cohomology classes count both the charged matter zero modes and instanton zero modes in case one of the seven-branes is a Euclidean three-brane instanton.

In the string theory literature, some indications how such a generalization of Bott's theorem to general toric spaces might look like implicitly appeared in the papers \cite{Distler:1996tj} and \cite{Blumenhagen:1997vt}, however to our knowledge the algorithm used in those papers was never really completely revealed.
\cite{Note1}
More recently, it was in the mathematics literature that the problem was systematically approached and solved \cite{CoxLittleSchenk:ToricVarieties}. However, it appears to us that the algorithm used in \cite{Distler:1996tj, Blumenhagen:1997vt} is much easier and economical, so that now, after non-trivially extending it to its most generic form, we find it appropriate to eventually reveal it. 

In this letter we will proceed by motivating the algorithm by a couple of non-trivial examples based on direct computations of the \v{C}ech cohomology. Then the algorithm is formulated as a conjecture (which is so simple that it fits less than a single page). We postpone the computation of vector bundle cohomologies and many further string theoretic applications to a more extensive paper \cite{PaperSecondPart}.
\\
\\
The high-performance {C/\CPP} implementation {\tt cohomCalg} of the algorithm is available under
\[
  \text{\href{http://wwwth.mppmu.mpg.de/members/blumenha/cohomcalg/}{\shortstack[r]{http://wwwth.mppmu.mpg.de/members/...\\...blumenha/cohomcalg/}}.}
\]

\section{Algorithm for line bundle cohomology}

In this section we will present an algorithm for the computation of line bundle cohomologies $H^i(X;\cO(D))$ over a toric variety  $X$. 
We will motivate the algorithm following the original development. The final conjecture is given in section \ref{sec:finalconjecture}.

\subsection{Preliminaries}

To define the framework we are working in, let us briefly summarize some facts about toric varieties respectively,  their physical counter part,  the gauged linear sigma models (GLSM).

A toric variety is a generalization of projective spaces, which contains a set of homogeneous coordinates $H\ce\{x_i : i=1,\ldots, I\}$ equipped with a number $R$ of equivalence relations
\beq
 (x_1,\dots,x_I)\sim (\lambda_r^{Q^{(r)}_1} x_1, \ldots, \lambda_r^{Q^{(r)}_I} x_I), \\
\eeq
for $r=1,\dots, R$ with the weights $Q^{(r)}_i\in \mathbb Z$ and $\lambda_r\in \IC^* = \IC-\{0\}$. We will always assume that the weights are chosen such that one really gets a bona fide toric space.  A powerful method to study such spaces is the GLSM \cite{Witten:1993yc}, in which the homogeneous coordinates are chiral superfields and the $\smash{Q^{(r)}_i}$ are charges under $R$ Abelian $U(1)$ gauge symmetries in a two-dimensional ${\cal N}=(2,2)$ supersymmetric gauge theory. The Fayet-Iliopoulos parameters $\xi_r$ of these $U(1)$ gauge symmetries, can be interpreted as the K\"ahler parameters of the geometric space.

The vanishing of the D-terms of a GLSM then splits the space of Fayet-Iliopoulos parameters $\vec \xi\in \IR^{R}$ into $R$-dimensional cones, in which the D-flatness conditions can be solved. These cones are  also called phases and correspond  geometrically to the  K\"ahler cones. In each such cone, for the D-terms to be solvable one finds sets of  collections of coordinates
\beq
	\cS_\alpha = \big\{ x_{\alpha_1}, x_{\alpha_2}, \ldots, x_{\alpha_{\vert\cS_\alpha \vert}} \big\} \quad  \textrm{with $\alpha=1,\ldots, N$},
\eeq  
which are not allowed to simultaneously vanish. The toric variety of dimension $D=I-R$ in this phase is then defined as
\beq
	X=\frac{\IC^I-F}{(\IC^*)^R } 
\eeq 
with the extracted set
\beq
	F=\bigcup_{\alpha=1}^N  \{x_{\alpha_1}=x_{\alpha_2}= \ldots=x_{\alpha_{\vert \cS_\alpha \vert}}=0\}\; .
\eeq
The information contained in  $F$ is nothing else than the Stanley-Reisner ideal ${\rm SR}$ of the toric variety $X$
\beq
  {\rm SR} = \left\langle {\cal S}_1,{\cal S}_2, \ldots, {\cal S}_N \right\rangle\; .
\eeq
Note that the more physical definition of a toric variety given here is equivalent to the one in terms of cones, fans and triangulations. A choice of a phase corresponds to the choice of a  triangulation of the polytope defined via the $I$ vertices $\nu_i\in \mathbb Z^D$. These vertices satisfy the $R$ linear relations
\beq
  \sum_{i=1}^I Q_i^{(r)}\nu_i = 0 \qquad \text{for } r=1,\dots,R.
\eeq
For further introductions to toric geometry consult \cite{Denef:2008wq, Kreuzer:2006ax, Reffert:2007im, HoriEA:MirrorSymmetry, Aspinwall:1993nu} or for a more mathematical treatment \cite{Fulton:ToricVarieties, Cox:ToricVarieties}.

Given such a toric space, the hypersurface $\{x_i=0\}$ naturally defines a divisor $D_i$ of $X$.  By the relation of divisors and line bundles summarized in appendix \ref{appline}, this also defines a line bundle $L_i={\cal O}_X(D_i)$, which we also denote as
\beq
	L_i={\cal O}_X \bigl(Q^{(1)}_i, \ldots, Q^{(R)}_i \bigr)\; .
\eeq
The tangent bundle $T_X$ of the toric variety is defined as the cokernel of the map $\beta$ in  the sequence
\beq\label{eq_generalizedeuler}
	0 \fto \cO_X^{\oplus R} \stackrel{\alpha}{\injto}  \bigoplus_{i=1}^I \cO_X(D_i) \stackrel{\beta}{\surjto} T_X \fto 0 \; . 
\eeq
In order to compute, for instance, the vector bundle cohomology $H^q(X;T_X)$, one needs as input only the line bundle cohomology classes $H^q(X;  \cO_X(D_i))$ with $q=0,\ldots,D$.

The determination of these cohomology classes for general line bundles $L={\cal O}_X(k_1, \ldots, k_R)$ is the problem we will address in this letter. Due to the Riemann-Roch-Hirzebruch theorem, the cohomology classes $H^q(X;L)$ satisfy an index theorem (see appendix \ref{appindex})
\beq
	\chi(X;L) = \sum_{q=0}^{D} (-1)^q\;  h^{q}(X;L) = \int_X {\rm ch}(L)\;  {\rm Td}(X)\; ,
\eeq   
where the right hand side can be computed easily, once the intersection form on $X$ is known.

The simplest toric varieties are the projective spaces $\IP^n$, for which $F=\{x_1=x_2=\ldots =x_{n+1}=0\}$. For these spaces, the line bundle cohomology is known due to Bott's theorem
\beq
  h^q(\IP^n;\cO(k)) = \begin{cases} \binom{n+k}{n} & {\rm for}\ q=0,\ k\ge 0 \\
                                    \binom{-k-1}{n} & {\rm for}\ q=n,\ k< -n
                                    \\
                                    0 & {\rm else} 
                      \end{cases}\; .
\eeq
The goal is now to generalize this result to toric ambient spaces.

\subsection{\texorpdfstring{\v{C}ech cohomology for $\IP^2$}{Cech cohomology for P2}}

To motivate the algorithm, we consider a simple example for which the \v{C}ech cohomology can really be computed with  pencil and paper.
 
\begin{figure}[ht]
  \centering
  \includegraphics[width=8cm]{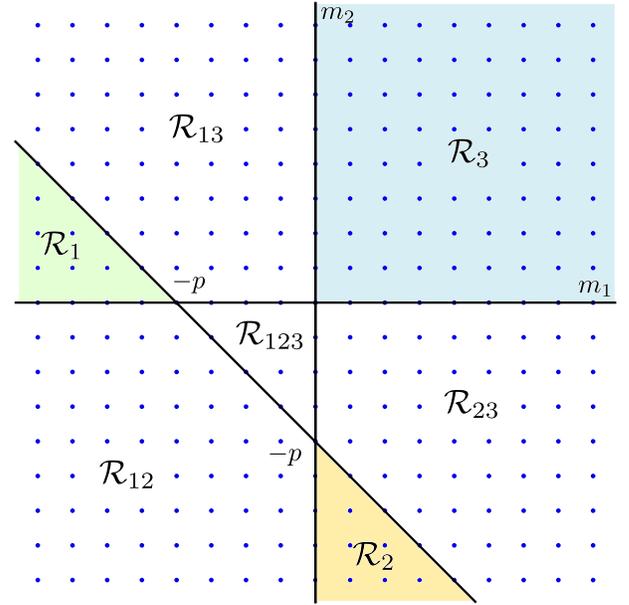}
  \caption{\small The region subdivision of the lattice. Note, that it is meant that we have the inclusions ${\cal R}_i\cup {\cal R}_j\subset {\cal R}_{ij}$ for $i<j$ and $\bigcup_i {\cal R}_i\cup \bigcup_{i<j} {\cal R}_{ij} \subset {\cal R}_{123}$, where ${\cal R}_{123}$ covers the entire lattice space.}
  \label{fig:regions}
\end{figure}

We want to compute the  \v{C}ech cohomology of $\IP^2$. First we need a suitable covering  by  open sets, which for this projective space is provided by
\beq
  U_i \ce \big\{ [x_1:x_2:x_3] \in \IP^2 : x_i \not= 0 \big\}\; .
\eeq
Then the \v{C}ech cochains for the covering $\fU_{\IP^2} = \{ U_1, U_2, U_3 \}$ take the form
\beq
  \bal
	  \check{C}^0(\fU_{\IP^2};\shF) & {} = \shF(U_1) \oplus \shF(U_2) \oplus \shF(U_3) \\
		\check{C}^1(\fU_{\IP^2};\shF) & {} = \shF(U_{12}) \oplus \shF(U_{23}) \oplus \shF(U_{13}) \\
		\check{C}^2(\fU_{\IP^2};\shF) & {} = \shF(U_{123}),
	\eal
\eeq
where we used the abbreviation $U_{ij\dots k} = U_i \cap U_j \cap \dots \cap U_k$. Now we consider a line bundle $\shF = \cO(-p)$ with $p>0$ and compute the local sections on all open sets, where the homogeneous coordinates which are non-vanishing can appear with a negative exponent. This leads to
\begin{itemize}
  \item $U_i$: an element of $\shF(U_i)$ is of the form
    \beq
		  \bal
				&x_1^{m_1}\, x_2^{m_2}\, x_3^{-p-m_1-m_2}, \\
				&{\rm with}\quad \begin{cases}  m_1,m_2\ge 0 & {\rm for\ } U_3 \\
				m_1\ge 0, \ m_2\le -p-m_1 & {\rm for\ } U_2\, . \\
				m_1\le -p-m_2,\  m_2\ge 0 & {\rm for\ } U_1 \\
				\end{cases} 
			\eal
    \eeq
    These monomials can be represented by points lying in different regions ${\cal R}_i$ in the $(m_1,m_2)$-plane, cf.~figure \ref{fig:regions}.
  \item $U_i\cap U_j$:\  an element of $\shF(U_i)\cap \shF(U_j)$ is of the form
		\beq
		  \bal
				& x_1^{m_1}\, x_2^{m_2}\, x_3^{-p-m_1-m_2}, \\
				& {\rm with}\quad \begin{cases}  m_2\ge 0 & {\rm for\ } U_1\cap U_3 \\
				m_1\ge 0  & {\rm for\ }
				U_2\cap U_3 \, . \\
				m_1+m_2\le -p & {\rm for\ }
				U_1\cap U_2 \\
				\end{cases} 
			\eal
		\eeq
    The corresponding regions are denoted by ${\cal R}_{ij}$ in the $(m_1,m_2)$-plane.
	\item $U_1\cap U_2\cap U_3$:\  an element of $\shF(U_1)\cap \shF(U_2)\cap \shF(U_3)$ is of the form
		\beq
			\bal
				& x_1^{m_1}\, x_2^{m_2}\, x_3^{-p-m_1-m_2}, \\	
				& {\rm with}\quad   (m_1,m_2)\in {\mathbb Z}^2 \; .
			\eal
		\eeq
    This means that ${\cal R}_{123}$ is the whole plane.
\end{itemize}

Now one has to distinguish the various regions in the $(m_1,m_2)$ plane shown in figure \ref{fig:regions}. Computing the cohomology of the \v{C}ech complex for those regions is straightforward. For $(m_1,m_2)\in {\cal R}_i$ each section of  $\check{C}^0(\fU_{\IP^2};\shF)$ restricts to two sets from  $\check{C}^1(\fU_{\IP^2};\shF)$, which by itself restricts to  $\check{C}^2(\fU_{\IP^2};\shF)$. Thus we get a sequence $0\to 1 \to 2 \to 1 \to 0$ with trivial cohomology. An element $(m_1,m_2)\in {\cal R}_{ij}\setminus {\cal R}_i\cup{\cal R}_j$ first contributes to $\check{C}^1(\fU_{\IP^2};\shF)$ and restricts to $\check{C}^2(\fU_{\IP^2};\shF)$. The induced sequence $0\to 0\to 1\to 1 \to 0$ again has trivial cohomology. Finally, if \beq (m_1,m_2)\in {\cal R}_{123}\setminus\bigcup_{i<j}{\cal R}_{ij}\cup\bigcup_{i}{\cal R}_i\, , \eeq  we get a sequence $0\to 0\to 0\to 1\to 0$  which contributes one element to the second cohomology.

Thus we have computed that the dimension of the cohomology $H^2(\IP^2;\cO(-p))$ for $p>0$  is given by the number of points lying in the interior triangle in the picture, i.e.~the complement of the union of all other sets. Now, we observe that the number of these points is precisely the number of all rational functions of the form 
\beq
	\frac{1}{x_1\, x_2\, x_3\, W(x_1,x_2,x_3)} \quad {\rm with} \quad \deg(W)=p-3\;,
\eeq
where $W$ is in fact only a monomial instead of a generic polynomial. From this point on we will refer to rational functions with both monomials in the numerator and denominator as ``rationoms''. This readily yields the result $h^2(\IP^2;\cO(-p))=\binom{p-1}{2}$, which is equal to Bott's formula.

The crucial observation is now that the combination $\{x_1 x_2 x_3\}$ is nothing else than the Stanley-Reisner ideal of $\IP^2$. Therefore, there seems to be a connection between such rationoms with the elements of the Stanley-Reisner ideal in the denominator and the dimensions of line bundle cohomology classes.

One can study more complicated examples and always finds a similar relationship between the representatives of \v{C}ech cohomology and rationoms of the above form. This observation leads to the following first version  of our conjecture, which does not yet cover the most generic situation. However, it was this form which was sufficient to find all the cohomology classes in \cite{Distler:1996tj, Blumenhagen:1997vt}.

\subsection{A preliminary algorithm conjecture}

For the moment we assume that all elements of the Stanley-Reisner ideal are disjoint, i.e.~no homogeneous coordinate $x_i$ appears in more than one ${\cal S}_\alpha$.

In order to determine which rationoms contribute to which cohomology, we have to consider the power set of ${\rm SR}$, which we denote as $P({\rm SR})$. Clearly we have a splitting of the form 
\beq
  P({\rm SR}) = \bigcup_{k=0}^{\left\vert {\rm SR} \right\vert} P_k({\rm SR}),
\eeq
where $P_k({\rm SR})$ contains all subsets of ${\rm SR}$ with $k$ elements.

Let $A = \lbrace \alpha_1, \dots , \alpha_k \rbrace \subset \lbrace 1, \dots , \left\vert {\rm SR }\right\vert \rbrace $ and $\cP^k_A=\{\cS_{\alpha_1}, \dots, \cS_{\alpha_k}\}\in P_k({\rm SR})$ be such an element of $P({\rm SR})$ containing $k$ elements. We now define the set
\beq
  \cQ^k_A \ce \bigcup_{i=1}^k\  \cS_{\alpha_i}
\eeq
of all homogeneous coordinates appearing in $\cP^k_A$ and the c-degree
\beq
  N^k_A\ce \left| \cQ_A^k \right| - k = \left\vert\, \bigcup_{i=1}^k\  \cS_{\alpha_i}\, \right\vert -k.
\eeq
The question now is to which cohomology class does a fixed element $\cQ=\cQ^k_A$ with $N=N^k_A$ contribute. 

\begin{description}
	\item[{\it Preliminary Conjecture:}] One gets a contribution to $H^{N}(X, \cO(D))$, which is given just by counting appropriate rationoms, where $\vec y \in \cQ$ are the denominator variables and $\vec x \in H-\cQ$ the complementary numerator variables.  The contribution of $\cQ$ to the line bundle cohomology is given by:
	\beq
		\boxed{H^{N}(X; \cO(D)): \qquad  \frac{T(\vec x)}{\left( \prod y_j\right)\cdot  W(\vec y)} }\; .
	\eeq
	$T(\vec x)$ and $W(\vec y)$ are monomials of the right degree to match the weights of $D$ in the line bundle $\cO(D)$, i.e.~one counts the number of suitable monomials $T$ and $W$. 
\end{description}

\noindent
Now, doing this for all elements in the power set $P({\rm SR})$, one determines the complete line bundle cohomology. The whole cohomology computation has been reduced to a counting problem, in which the Stanley-Reisner ideal plays an important role. Clearly, this algorithm  can easily  be computerized.

\subsection{Further examples}

To get an idea of how this algorithm works in more involved cases and where its restrictions are, we now apply this conjecture to del Pezzo surfaces arising from a certain number of blowups of $\IP^2$.

\subsubsection*{del Pezzo-1 surface}
To get an idea of the simplicity of the counting procedure once the form of the contributing monomials has been determined, we first look at $dP_1$ with toric data given in table \ref{tab:dPoneSurface}. Note that the condition on the elements of the Stanley-Reisner ideal is fulfilled, i.e.~they are disjoint. We will see what may happen for more complicated Stanley-Reisner ideals later.

\begin{table}[ht]
  \centering
  \begin{tabular}{r@{\,$=$\,(\,}r@{,\;\;}r@{\,)\;\;}|c|cc|c}
    \multicolumn{3}{c|}{vertices of the} & coords & \multicolumn{2}{c|}{GLSM charges} & {divisor class}${}^\big.$ \\
    \multicolumn{3}{c|}{polyhedron / fan} & & $Q^1$ & $Q^2$ & \\
    \hline\hline
    $\nu_1$ & $-1$ & $-1$ & $x_1$ & 1 & 0 & $H$\uspc \\
    $\nu_2$ &  1   &  0   & $x_2$ & 1 & 0 & $H$  \\
    $\nu_3$ &  0   &  1   & $x_3$ & 1 & 1 & $H+X$ \\
    $\nu_4$ &  0   & $-1$ & $x_4$ & 0 & 1 & $X$
  \end{tabular}
  \\[5mm] intersection form: ${}\quad HX - X^2$
  \\[3mm] ${\rm SR}(dP_1) = \langle x_1 x_2 ,\; x_3 x_4 \rangle = \langle \cS_1, \cS_2 \rangle$
  \caption{\small Toric data for the del Pezzo-1 surface}
  \label{tab:dPoneSurface}
\end{table}

Forming the power set of the Stanley-Reisner ideal directly yields the following possible contributions to cohomology, where the line bundle $\cO (D)= \cO(m,n)$ is given by the divisor $D=mH+nX$.  
\beq
	\bal
		H^0(dP_1; \cO(m,n)):& \qquad T(x_1, x_2, x_3, x_4)\, ,\\
		H^1(dP_1; \cO(m,n)):& \qquad \frac{T(x_3,x_4)}{x_1 x_2\cdot W(x_1,x_2)}\, , \\
												& \qquad \frac{T(x_1,x_2)}{x_3 x_4 \cdot W(x_3,x_4)}\, ,\\
		H^2(dP_1; \cO(m,n)):& \qquad \frac{1}{x_1 x_2 x_3 x_4 \cdot W(x_1,x_2,x_3,x_4)}\, .  
	\eal 
\eeq

In fact, we would not have to take $H^2(dP_1; \cO(m,n))$ into account because of Serre duality, but it will serve as a good test of the procedure to check that it is obeyed at the end of the computation. Note that the rationoms representing the top cohomology class always involve the denominator monomial $\cQ=\prod x_i$ of all homogeneous coordinates $x_i\in H$, which is a clear evidence of Serre duality in ``monomial'' form, since the canonical class $K=-\sum D_i$ is the negative sum of all coordinate divisor classes.

In order to get the dimensions of all cohomology groups, one only has to read off the charges of the coordinates from the table to get the degrees of the respective monomials, equate them to the degrees $(m,n)$ of the line bundle and do the remaining combinatorics, i.e.~count all possible exponents. Writing $h^i(m,n)$ for the dimension of $H^i(dP_1; \cO (m,n))$ and $\labs x_i \rabs$ for the exponent of the coordinate $x_i$ in the monomials $T$ and $W$ these steps can be schematically described as follows: 
\begin{itemize}
  \item Contributions to $h^0(m,n)$ from all combinations of exponents with
    \beq
			\bal
      & \deg \,\, T(x_1, x_2, x_3, x_4) \\
			& {} = \left( \labs x_1 \rabs + \labs x_2 \rabs + \labs x_3 \rabs, \, \labs x_3 \rabs + \labs x_4 \rabs \right) \\
			& {} \stackrel{!}{=} (m, n)\, .
			\eal
    \eeq
    Since the exponents cannot be negative, we only get contributions for $m\geq 0$ and $n\geq 0$. Defining $\binom{r}{2}=0$ if $r<2$, one calculates
    \beq
    h^0(m,n)=\binom{m+2}{2}-\binom{m-n+1}{2}\, .
    \eeq
  \item The degrees of the monomials contributing to $h^1(m,n)$ can be evaluated as 
    \beq 
      \bal
        &\deg \,\, \frac{T(x_3,x_4)}{x_1 x_2\cdot W(x_1,x_2)} \\
				& {} = \left( \labs x_3 \rabs -2 -\labs x_1 \rabs -\labs x_2 \rabs ,\, \labs x_3 \rabs + \labs x_4 \rabs \right)\, ,\\
        & \deg \,\, \frac{T(x_1,x_2)}{x_3 x_4\cdot W(x_3,x_4)} \\
				& \mrlap{{} = \left( \labs x_1 \rabs + \labs x_2 \rabs -1 -\labs x_3 \rabs ,\, -2 - \labs x_3 \rabs - \labs x_4 \rabs \right).  }
      \eal 
    \eeq 
    Equating the right hand sides to $(m,n)$, this yields the two cases
    \begin{enumerate}
      \item $n\geq 0\,  \wedge \, n\geq m+2$: \\ 
			      $\qquad\! h^1(m,n)= \binom{n-m}{2}-\binom{-m-1}{2}\, .$
      \item $n\leq -2\, \wedge\, n\leq m-1$:  \\
			      $\quad h^1(m,n)  = \binom{m-n+1}{2}-\binom{m+2}{2}\, .$
    \end{enumerate}
    This again makes use of the convention $\binom{r}{2}=0$ if $r<2$.
  \item Contributions to $h^2(m,n)$ come from a monomial with degree 
    \beq
      \bal
        &\deg \frac{1}{x_1 x_2 x_3 x_4\cdot W(x_1,x_2,x_3,x_4)} \\
        &{} =(-3 - \labs x_1 \rabs - \labs x_2 \rabs - \labs x_3 \rabs, \\
				&{}\qquad{}-2 - \labs x_3 \rabs - \labs x_4 \rabs )\, .
      \eal
    \eeq
    Setting this equal to $(m,n)$, we only get contributions when $m\leq 0$ and $n\leq 0$. The result is
    \beq
    h^2(m,n)=\binom{-m-1}{2}-\binom{n-m}{2}\, .
    \eeq
\end{itemize}

It is easy to check that Serre duality holds. Since the canonical divisor is given by the negative sum over all divisors corresponding to the one-dimensional cones (vertices) of the fan 
\beq 
  K=-\sum_{\mclap{\rho\in \Sigma(1)}}D_\rho\, , 
\eeq 
we get $K= -3H -2X$ from the table of $dP_1$. Serre duality can then be written as 
\beq 
  \bal
    & H^i(dP_1;\cO(m,n)) \\
		&{}\cong H^{2-i}(dP_1;\cO(K)\otimes \cO(m,n)^{\vee}) \\
    &{}\cong H^{2-i}(dP_1 ;\cO(-m-3,-n-2))\, .  
	\eal 
\eeq 
And indeed, the computed dimensions obviously satisfy the identity 
\beq
  h^i(m,n)=h^{2-i}(-m-3,-n-2)\, .  
\eeq 
In this case the computation could still be performed just with pencil and paper but clearly for more involved higher dimensional cases a computer code is necessary.
\cite{Note2}

\subsubsection*{del Pezzo-3 surface}
So let's make two steps further and look at $dP_3$, the del Pezzo surface of degree 6 coming from three consecutive blowups of $\IP^2$. This is the first example we considered, for which nontrivial ``region factors'' arise from \v{C}ech cohomology. Its toric data are given in Table \ref{tbl:dPthreeSurface}.

\begin{table}[ht]
  \begin{center}
    \begin{tabular}{r@{\,$=$\,(\,}r@{,\;\;}r@{\,)\;\;}|c|cccc|c} 
      \multicolumn{3}{c|}{vertices of the} & coords & \multicolumn{4}{c|}{GLSM charges} & {divisor class}${}^\big.$ \\
      \multicolumn{3}{c|}{polyhedron / fan}      &        & $Q^1$ & $Q^2$ & $Q^3$ & $Q^4$ & \\ 
			\hline\hline
      $\nu_1$ & $-1$ & $-1$ & $x_1$ & 1 & 0 & 0 & 1 & $H+Z$\uspc \\
      $\nu_2$ &   1  &   0  & $x_2$ & 1 & 0 & 1 & 0 & $H+Y$ \\
      $\nu_3$ &   0  &   1  & $x_3$ & 1 & 1 & 0 & 0 & $H+X$ \\
      $\nu_4$ &   0  & $-1$ & $x_4$ & 0 & 1 & 0 & 0 & $X$ \\
      $\nu_5$ & $-1$ &   0  & $x_5$ & 0 & 0 & 1 & 0 & $Y$ \\
			$\nu_6$ &   1  &   1  & $x_6$ & 0 & 0 & 0 & 1 & $Z$ 
    \end{tabular}
    \\[5mm] intersection form: ${}\quad HX + HY + HZ - 2H^2 - X^2 - Y^2 - Z^2$
		\\[3mm] ${\rm SR}(dP_3) = \langle \underbrace{x_1 x_2}_{\cS_1},\; \underbrace{x_1 x_3}_{\cS_2}, \; \underbrace{x_1 x_6}_{\cS_3}, \; \underbrace{x_2 x_3}_{\cS_4}, \; x_2 x_5, \; x_3 x_4, \; x_4 x_5, \; x_4 x_6, \; \underbrace{x_5 x_6}_{\cS_9} \rangle$
    \vspace{-3mm}
  \end{center}
  \caption{\small Toric data for the del Pezzo-3 surface.}
  \label{tbl:dPthreeSurface}
\end{table}

First of all, the elements of the Stanley-Reisner ideal are no longer disjoint. So by our definition of c-degree, e.g.~the element $\cP_{\lbrace 1,4 \rbrace}^2=\lbrace x_1 x_2 ,x_2 x_3 \rbrace$ of $P_2({\rm SR})$ would now contribute to the first cohomology, since the union is 
\beq
  \cQ\ce\cQ^2_{\lbrace 1,4 \rbrace }=\lbrace x_1, x_2, x_3\rbrace
\eeq
and therefore $N^2_{\lbrace 1,4 \rbrace }=3-2=1$. There are also two other combinations in $P_2({\rm SR})$ with the same union, namely $\cP^2_{\lbrace 2,4 \rbrace }=\lbrace x_1 x_3,x_2 x_3 \rbrace$ and $\cP^2_{\lbrace 1,2 \rbrace }=\lbrace x_1 x_2 ,x_1 x_3 \rbrace$. This by itself is not a problem, but one further finds out that the c-degree of $\cQ$ is not unique any more, since the element $\cP^3_{\lbrace 1,2,4 \rbrace }=\lbrace x_1 x_2 ,x_1 x_3,x_2 x_3 \rbrace$ of $P_3({\rm SR})$ has the same union but now gives $N^3_{\lbrace 1,2,4 \rbrace }=3-3=0$.  Of course one could hope that this kind of combinatorial structure of the power set is meaningless and a certain element $\cQ$ only contributes to the cohomology of highest c-degree.

As it turns out, this assumption is too naive, since it leads to inconsistencies for example when computing the holomorphic Euler characteristic of line bundles, which can also be computed by means of index theorems and is another useful quantity to cross-check results. Choosing $\cO(D)$ with $D=-3H-X-Y-Z$ (equal to the degree of $x_1^{-1} x_2^{-1} x_3^{-1}$), computing the Todd class of $dP_3$, the Chern character of the line bundle and doing the integral over $X$ by means of the intersection form, one readily calculates
\beq
  \bal
		& \chi(dP_3;\cO(D))\\
		& {} =h^0(\cO(D))-h^1(\cO(D))+h^2(\cO(D))=-2
	\eal
\eeq
and since $h^0(\cO(D))=h^2(\cO(D))=0$ we get $h^1(\cO(D))=2$. But performing our algorithm, the only contribution comes from the rationom $x_1^{-1} x_2^{-1} x_3^{-1}$, so it clearly fails. The rationom must be weighted by a factor of two and the question is how this can be incorporated in the counting algorithm.

A possible solution is to introduce a sort of ``cohomological'' sequence that imitates a cochain complex without specifying the mappings and encodes the additional structure of $P({\rm SR})$ alluded to before. In the case of $\cQ$ this yields
\beq
  \ldots \fto 0 \fto \underbrace{\fC^0}_{=1} \fto \underbrace{\fC^{1}}_{=3} \fto 0 \fto \ldots .
\eeq
and therefore a remnant or secondary cohomology $\fh^1=3-1=2$ acting as a factor for all corresponding rationoms, allowing us to schematically write the contribution of $\cQ$ to line bundle cohomology as
\beq
   H^{1}(dP_3; \cO(D)): \quad  2\, \times\, \frac{T(x_4,x_5,x_6)}{x_1 x_2 x_3 \cdot W(x_1,x_2,x_3)}\; ,   
\eeq
where the monomials $T$ and $W$ have to be chosen such that the rationom has the same degree as $D$. Since it can be shown that this procedure is suited to reproduce the right factors in many more nontrivial examples including higher-dimensional varieties, this motivates the formulation of a more general version of our conjectured algorithm.

\subsection{The final algorithm conjecture}\label{sec:finalconjecture}
To formulate it we have to refine the set-theoretic notions from the first version of the conjecture.  In order to determine which rationoms contribute to which cohomology, we have to consider the power set of ${\rm SR}$, which we denote as $P({\rm SR})$. Clearly we have a splitting,  \beq P({\rm SR}) = \bigcup_{k=0}^{\left\vert {\rm SR} \right\vert} P_k({\rm SR})\; , \eeq where $P_k({\rm SR})$ contains all subsets of ${\rm SR}$ with $k$ elements.  Let $A = \lbrace \alpha_1, \dots , \alpha_k \rbrace \subset \lbrace 1, \dots , \left\vert {\rm SR }\right\vert \rbrace $ and $\cP^k_A=\{\cS_{\alpha_1}, \dots, \cS_{\alpha_k}\}$ be such an element of $P({\rm SR})$ containing $k$ elements. We now define the set
\beq
  \cQ^k_A \ce \bigcup_{i=1}^k\ \cS_{\alpha_i} 
\eeq
of all homogeneous coordinates appearing in $\cP^k_A$ and the c-degree 
\beq
  N^k_A\ce \left\vert \cQ_A^k \right\vert - k = \left\vert\, \bigcup_{i=1}^k\ \cS_{\alpha_i}\, \right\vert -k\; . 
\eeq 
Now, for fixed $k$ we take the disjoint union 
\beq 
  \cQ^k \ce \coprod_{A} \cQ^k_A, 
\eeq 
where $A$ runs over all subsets of $\lbrace 1, \dots , \left\vert {\rm SR }\right\vert \rbrace$ with $k$ elements. The question now is to which cohomology class does a fixed denominator monomial $\cQ$ contribute. To decide this one computes a ``remnant'' or ``secondary'' cohomology $\cH^i(\cQ)$. For its definition, one counts how often the element $\cQ$ appears in each $\cQ^k$ and with what c-degree. Then we can associate the numbers $\fC^{i}(\cQ)$ to such a $\cQ$, where $\fC^i(\cQ)$ is the number of times the element $\cQ$ appears in all $\cQ^k$ with c-degree $i$.
\begin{description}
	\item[{\it Algorithm Conjecture:}] For each non-vanishing cohomology $\fh^i(\cQ)$ of the complex
		\beq
		\label{secseq}
			\ldots \fto \fC^0(\cQ) \fto \ldots \fto \fC^{d}(\cQ) \fto \dots\; 
		\eeq
		one gets a contribution to $H^{i}(X; \cO(D))$, which is given by counting rationoms
		\beq
			\boxed{H^{i}(X; \cO(D)): \quad  \fh^i(\cQ)\, \times\, \frac{T(\vec x)}{\left( \prod y_j\right)\cdot  W(\vec y)}}\; ,
		\eeq
		where $\vec y\in \cQ$ are  the denominator variables and $\vec x \in H-\cQ$ the numerator variables. $T(\vec x)$ and $W(\vec y)$ have to be taken as monomials of the right degree to match the degree of $\cO(D)$. The total number of these has then to be weighted by a factor of $\fh^i(\cQ)$.
\end{description}

\bigskip\noindent
Actually, due to Serre duality it is sufficient to consider only the cohomologies $H^i(X; \cO(D))$ with $i\in\{0,1,\ldots, [\frac{d}{2}]\}$. We have not really been precise about what the maps in \eqref{secseq} are and therefore have not shown that  this is indeed a complex. More mathematical work needs to be done to precisely define this ``remnant'' cohomology. This would mean to see in detail that, after taking the localization of the line bundle cohomologies onto the elements of $P({\rm SR})$ into account, one has only computed the complete  \v{C}ech cohomology up to this remnant sequence.

We have tested the conjecture in many situations, with the most complicated one being line bundles on a Calabi-Yau fourfold with a resolved $SU(5)$ singularity of a certain divisor. Here the Stanley-Reisner ideal involved about 20 generators with many coordinates appearing more than once. 

\section{Summary and Conclusions}
Based on earlier  observations in  \cite{Distler:1996tj} and \cite{Blumenhagen:1997vt}, in this letter we presented a general working algorithm for the computation of line bundle valued cohomology classes over D-dimensional toric varieties. Evaluating  the  \v{C}ech cohomology for certain comparably simple examples, we realized that the representatives of the local sections generating $H^i(X;L)$ with $i=0,\ldots, D$ are given by certain rational functions in the homogeneous coordinates, where the elements of the Stanley-Reisner ideal appeared in the denominator. The algorithm became more involved by realizing that for the case that homogeneous coordinates appeared more than once in the Stanley-Reisner ideal, extra multiplicities had to appear. For the determination of these extra factors, we proposed to evaluate a remnant cohomology, which we suspect  can be thought of as the not yet fully resolved  part  of the original \v{C}ech cohomology, after the localization on the Stanley-Reisner ideal sets has been done.

We checked many examples and 
found no deviation from known results, Serre duality or the Riemann-Roch-Hirzebruch index theorem. 
The algorithm also is consistent with results obtained by the algorithm presented recently in \cite{Cox:ToricVarieties}. Clearly, the advantage of the latter is that it has been mathematically proven, but we want to emphasize that, in the cases we tested, our algorithm was much more efficient. Indeed, examples that need about 10 minutes of computing time using the ``chamber'' algorithm could still be done with pencil and paper using the algorithm presented here. 

Finally, having available such a simple and easy to implement algorithm paves the way to perform a whole amount of concrete computations in string theory. This includes heterotic string compactifications over Calabi-Yau manifolds as well as intersecting D-brane models and F-theory compactifications on Calabi-Yau fourfolds. We will report on the applications to these string theoretic cases in a future publication \cite{PaperSecondPart}.


\subsubsection*{Added: Proof of the algorithm}
About three months after the preprint release of this work, a proof of the conjectured algorithm was presented in \cite{Roschy:2010fm}, which also clarifies much of the underlying mathematical structures. At the same time an independant proof was developed in \cite{2010arXiv1006.0780J} --- published in fact a few days earlier --- which utilized alternative methods.

\subsection*{Acknowledgement}
We gratefully thank our computers for doing most of the work for us and R.~Blumenhagen acknowledges a brief conversation with J.~Distler in 1997. Furthermore, we would like to thank A.~Collinucci for a number of useful comments on the manuscript. R.~Blumenhagen and B.~Jurke would like to thank the Kavli Institute for Theoretical Physics, Santa Barbara for the hospitality during the final stages of the project. This research was supported in part by the National Science Foundation under Grant No. PHY05-51164.

\vspace{1cm}
\noindent
Note added: During the very final stages of this work, we became aware of \cite{Cvetic:2010rq} which partially addresses the same problem.

\appendix

\section{Mathematical Background}

This section serves as a reference to the mathematical notions and structures used later on. We recall the definition of sheaves, which are abundantly used in mathematical physics in the context of vector bundles, and summarize the basic ideas behind sheaf and \v{C}ech cohomology. The language is therefore much more formal compared to the first part of this paper.

\subsection{Sheaves}

In mathematics one often encounters spaces which are defined by stitching together local patches in a suitable fashion. Therefore it seems natural to attach additional data directly to those patches, i.e.~to local subsets of the described total space. This idea is formalized using the notion of a sheaf, which can also be treated as a generalization of the ordinary bundle concept, where the fiber space attached to every point of the base space may vary.

Let us recall the formal definition of a sheaf:
\cite{Note3}
Given a topological space $X$, a sheaf $\shF$ of Abelian groups or modules on $X$ assigns to each open subset $U\subset X$ an Abelian group or module $\shF(U)$, such that
\begin{enumerate}
	\item \emph{Normalization}: $\shF(\emptyset) = 1$, the trivial group or module.
	\item \emph{Local uniqueness}: If $\fU=\{U_i\}$ is an open covering of $U \subset X$ and $s,t\in\shF(U)$ are two elements (``sections over $U$'') that are identical on each local subset, i.e.~$s|_{U_i} = t|_{U_i}$ for all $U_i\in\fU$, then in fact $s=t$ follows.
	\item \emph{Gluing}: If $U,V\subset X$ are two open sets and $s_U\in\shF(U)$ as well as $s_V\in\shF(V)$ two elements (``sections'') that are identical on the intersection, i.e.~$s_U|_{U\cap V} = s_V|_{U\cap V}$, then there exists a glued element $s\in\shF(U\cup V)$ such that $s|_U = s_U$ and $s|_V = s_V$.
\end{enumerate}
Note that for some subset $U\subset X$ the restriction $\shF|_U$ is itself a sheaf on $U$, given by $\shF|_U(V) = \shF(U\cap V)$, whereas $\shF(U)$ refers to the Abelian group or module attached to the subset.

In order to see that a sheaf generalizes the notion of bundles, let $E\stackrel{\pi}{\surjto} X$ be a vector bundle. For $U\subset X$ let $\Gamma(U,E)$ denote the space of sections over $U$, which naturally has a module structure by addition of sections, i.e.~fiberwise addition of the respective vectors, and multiplication with holomorphic functions. If we define $\shF$ via $\shF(U) \ce \Gamma(U,E)$ it can be shown that all the properties of a sheaf are satisfied. Therefore we observe that the formal transition from bundles to sheaves is essentially nothing else as treating them in terms of the local sections. In this sense the notation $\Gamma(U,\shF)$ is frequently used to refer to the elements of $\shF(U)$.

\subsection{Sheaf cohomology}

Given a sheaf $\shF$ on a variety $X$ one can define sheaf cohomology groups $H^p(X;\shF)$, which are similar to e.g.~de Rham cohomology groups of differential forms with values in a vector bundle. If
\beq
  0 \fto \shF \injto \shI^0 \stackrel{d^0}{\fto} \shI^1 \stackrel{d^1}{\fto} \shI^2 \stackrel{d^2}{\fto} \dots
\eeq
is an injective resolution of the sheaf $\shF$, denoted by $(\shI^\bullet,d^\bullet)$, there is an induced complex of global sections
\beq
  0 \fto \shI^0(X) \stackrel{\tilde d^0}{\fto} \shI^1(X) \stackrel{\tilde d^1}{\fto} \shI^2(X) \stackrel{\tilde d^2}{\fto} \dots
\eeq
satisfying $\tilde d^{p+1} \circ \tilde d^p = 0$ for all $p\ge 0$. Note that this only implies ${\rm im}(\tilde d^p) \subseteq {\rm ker}(\tilde d^{p+1})$, whereas for exactness of the sequence the equality must be strictly satisfied. The first term $\shF(X)$ of the induced complex $\smash{\big(\shI^\bullet(X),\tilde d^\bullet\big)}$ is usually omitted for notational convenience and $\smash{\tilde d^{-1}:0\injto \shI^0(X)}$ is the zero mapping. The sheaf cohomology groups of $\shF$ are then defined by
\beq
  H^p(X;\shF) \ce H^p\big(\shI^\bullet(X)\big) = \frac{{\rm ker}(\tilde d^p)}{{\rm im}(\tilde d^{p-1})}.
\eeq
One certainly has to verify a number of mathematical aspects in order to prove that this is well-defined, see \cite{Bredon:SheafTheory}.

As one may anticipate from the notion of cohomology, a sheaf homomorphism $\shF\fto\shG$ induces a corresponding homomorphism of sheaf cohomology groups $H^p(X;\shF)\fto H^p(X;\shG)$ and a short exact sequence of sheaves gives rise to a long exact sequence of sheaf cohomology groups. Furthermore, from the definition it follows that $H^0(X;\shF)=\shF(X)=\Gamma(X,\shF)$, i.e.~the 0th sheaf cohomology group can be treated as the space of global sections of the sheaf $\shF$ over $X$, which in our cases usually corresponds to global sections of the vector bundle described by the sheaf.

\subsection{\texorpdfstring{\v{C}}{C}ech cohomology}

The elegant abstract definition of sheaf cohomology is unfortunately rather unsuitable for actual computations. However, there is a another cohomology theory based on intersections of open sets covering $X$ that yields an equivalent cohomology, called \v{C}ech cohomology.

Let $\fU_X = \{ U_i \}_{i=1}^m$ be an open covering of the variety $X$, i.e.~$X=\bigcup_{i=1}^m U_i$ and every $U_i\subset X$ is an open set. As usual $\shF(U_i)$ refers to the Abelian group or module that the sheaf $\shF$ associates to the open set $U_i$. The \v{C}ech cochains are defined by
\beq
  \check{C}^p(\fU;\shF) = \bigoplus_{{}\quad\mclap{0\le i_0 < i_1 < \dots < i_p \le m}\quad{}} \shF(U_{i_0} \cap U_{i_1} \cap \dots \cap U_{i_p}),
\eeq
which means we are considering all the Abelian groups or modules, the sheaf $\shF$ associated to mutual intersections of $p+1$ open sets of the covering $\fU$. Let $\alpha=\alpha(i_0,\dots,i_p)$ denote an element of $\shF(U_{i_0} \cap \dots \cap U_{i_p})$. The exterior derivative --- i.e.~mapping of the complex --- is then defined by
\beq
  \bal
	  &\check{d}^p: \check{C}^p(\fU;\shF) \fto \check{C}^{p+1}(\fU;\shF)\, , \qquad \text{where} \\
		& \big(\check{d}^p(\alpha)\big)(i_0,\dots,i_{p+1}) \\
		& {} =\sum_{k=0}^{p+1} (-1)^k \alpha(i_0,\dots,\widehat{i_k},\dots,i_{p+1}) \big|_{U_{i_0}\cap\dots\cap U_{i_{p+1}}},
	\eal
\eeq
with the ``hatted'' index $\widehat{i_k}$ being omitted. Basically, we are considering the alternating sum with one index omitted, which is familiar from the exterior derivative of e.g.~the de Rham complex. Again, one can check the property $\check{d}^{p+1}\circ\check{d}^p = 0$, such that
\beq
  0 \fto \check{C}^0(\fU;\shF) \stackrel{\check{d}^0}{\fto} \check{C}^1(\fU;\shF) \stackrel{\check{d}^1}{\fto} \dots
\eeq
indeed gives the \v{C}ech complex $\big(\check{C}^\bullet(\fU;\shF),\check{d}^\bullet\big)$. The $p$th \v{C}ech cohomology group is then defined by
\beq
 \check{H}^p(\fU;\shF) = H^p\big(\check{C}^\bullet(\fU;\shF)\big) = \frac{{\rm ker}(\check{d}^p)}{{\rm im}(\check{d}^{p-1})},
\eeq
which as before measures the failure of the complex to be exact. Again, one has to prove several mathematical aspects in order to show that any potential ambiguities are indeed taken care of in this definition. Like for sheaf cohomology one can show that a sheaf homomorphism $\shF\fto\shG$ induces a mapping of \v{C}ech cohomology groups $\check{H}^p(\fU;\shF)\fto\check{H}^{p+1}(\fU;\shG)$ and a short exact sequence of sheaves yields a long exact sequence of \v{C}ech cohomology groups. Furthermore, it follows $\check{H}^0(\fU;\shF)=\shF(X)=\Gamma(X;\shF)$ which establishes the equivalence of both sheaf and \v{C}ech cohomology at the 0th level.

Naturally the question arises how the sheaf and \v{C}ech cohomology groups for $p>0$ are related. Let $\fU=\{U_i\}_{i=1}^m$ and $\fV=\{V_j\}_{j=1}^n$ be two open coverings of the variety $X$. Then $\fV$ is called a refinement of $\fU$ if there exists an index mapping $\phi$, such that $V_j \subseteq U_{\phi(j)}$ holds for all $j=1,\dots,n$ --- basically every open set of the covering $\fU$ may be broken down into several open subsets which provide the covering $\fV$. This establishes a partial ordering ``$\le$'' between different coverings and therefore allows to use the notion of direct limit. Then it can be proved that in general
\beq
  \underbrace{H^p(X;\shF)}_{\mclap{\text{sheaf cohomology}}} \cong \indlim\limits_{\fU} \underbrace{\check{H}^p(\fU;\shF)}_{\mclap{\text{\v{C}ech cohomology}}}
\eeq
holds for all $p\ge 0$, which basically implies that for a sufficiently refined open covering of $X$ both sheaf and \v{C}ech cohomology groups are identical. For the cases relevant to physicists the statement can be substantially improved: Let $\shF$ be a quasicoherent sheaf and $\fU=\{U_i\}_{i=1}^m$ an affine open covering of the variety $X$, then for all $p\ge 0$ there are natural isomorphisms
\beq\label{eq_sheafcechiso}
  H^p(X;\shF) \cong \check{H}^p(\fU;\shF).
\eeq

Essentially, those statements allow physicists to talk about sheaf cohomology but carry out all the actual computations using \v{C}ech cohomology. The huge advantage of \v{C}ech cohomology lies in the fact that for a suitable choice of the open covering $\fU$ the necessary explicit computations of the \v{C}ech complex are greatly simplified.

\subsection{Holomorphic line bundles and divisors}\label{appline}
The primary sheaf of interest in subsequent dealings is the sheaf of holomorphic functions on $X$, denoted by $\cO_X$. For each open set $U\subset X$ the corresponding module $\cO_X(U)$ is the set of holomorphic functions $U\fto\IC$. In fact, $\cO_X$ has a natural (commutative) ring structure induced by complex multiplication. In this context $\cO_X$ is called the structure sheaf of $X$ and turns it into a ringed space. Furthermore, there are the sheaves $\cO^*_X$ of nowhere zero holomorphic functions, $\cM_X$ of meromorphic functions and $\cM^*_X$ of non-trivial meromorphic functions, which are frequently used in algebraic geometry.

Those sheaves are naturally related to divisors, which are formal sums of irreducible hypersurfaces of $X$. The set of such Weil divisors is denoted ${\rm Div}(X)$ and it is in fact isomorphic to the set of Cartier divisors under a weak smoothness assumption on $X$, i.e.
\beq\label{eq_divcohomiso}
  {\rm Div}(X) \cong H^0(X;\cM^*_X / \cO^*_X).
\eeq
Now consider the set of isomorphism classes of holomorphic line bundles over $X$. The tensor product of vector bundles and the dual vector bundle naturally provide it with an Abelian group structure, which is called the Picard group ${\rm Pic}(X)$. Again, there is an isomorphism
\beq\label{eq_piccohomiso}
  {\rm Pic}(X) \cong H^1(X;\cO^*_X)
\eeq
relating to sheaf cohomology. Note that one usually ignores the fact that ${\rm Pic}(X)$ consists of the isomorphism classes of line bundles instead of the line bundles, which leads to a frequent abuse in notation. Now there exists a natural group homomorphism
\cite{Note4}
\beq\label{eq_naturaldivpicmap}
  \bal
	  \Xi:{}& {\rm Div}(X) \fto {\rm Pic}(X) \\
		&  D \mapsto \cO_X(D)
	\eal
\eeq
that maps some Weil divisor to a corresponding element of the Picard group, i.e.~it maps a divisor to an isomorphism class of holomorphic vector bundles. A corresponding vector bundle of this class can be defined as follows: 

Due to \eqref{eq_divcohomiso} any divisor corresponds to a global section $\rho$ of $\cM^*_X/\cO^*_X$, which can be specified in terms of functions $\rho_i\in\cM^*_X(U_i)$ for some covering $\fU=\{U_i\}$ of $X$. The line bundle representative corresponding to $\cO_X(D)$ is then specified by the transition functions $\smash{h_{ij}\ce\rho_i\cdot\rho_j^{-1}\in\cO^*_X(U_{ij})}$. It is necessary to show that the $h_{ij}$ indeed define an element of $\check{H}^1(\fU;\cO^*_X)$, such that $\cO_X(D)$ is a well-defined Picard generator via \eqref{eq_piccohomiso} and \eqref{eq_sheafcechiso}. Furthermore, $\cO_X(D)$ satisfies a number of useful properties:
\begin{itemize}
	\item \emph{Abelian group structure:} $\cO_X(D+D')=\cO_X(D)\otimes\cO_X(D')$
	\item \emph{Inverse elements:} $\cO_X(-D)\cong\cO_X(D)^*$
	\item \emph{Neutral element:} $\cO_X(0)\cong\cO_X$
	\item \emph{Naturality:} $\cO_X(f^*D)\cong f^*\cO_Y(D)$ for a holomorphic mapping $f:X\fto Y$.
\end{itemize}

\subsection{Index theorems}\label{appindex}

Due to the numerous sequences and steps one has to work through in order to derive the information of interest, it is useful to have a couple of non-trivial consistency checks at hand. Those are provided by index theorems, which allow to compute certain topological invariants via entirely different means. The Euler characteristic
\beq\label{eq_chirealdefinition}
  \chi(X) = \sum_{k=0}^{\mclap{\dim_\IR X}} (-1)^k b^k(X) = \int_X e(X)
\eeq
can be defined as either the alternating sum of the Betti numbers $b^i(X) \ce \dim_{\IR} H^i(X)$, or as the integral of a differential form representing the Euler class $e(X)$, which is equal to the top Chern class $c_n(X)\in H^{2n}(X)$ for some complex $n$-dimensional manifold.

For the K\"ahler manifolds that are primarily investigated via toric geometry, the prior statement can be generalized to a holomorphic version. Basically, the Euler class $e(X)$ is replaced by the Todd class ${\rm Td}(X)$, which is defined only for complex vector bundles. The Todd class can be expanded in terms of Chern classes like
\beq
  \bal
    {\rm Td} = {} & 1 + \frac{c_1}{2} + \frac{2c_1^2 + 2c_2 + c_1 c_2}{24} \\
		& {} + \frac{-c_1^4 + 4 c_1^2 c_2 + c_1 c_3 + 3 c_2^2 - c_4}{720} \\
		&{} + \frac{-c_1^3 c_2 + 3 c_1 c_2^2 + c_1^2 c_3 - c_1 c_4}{1440} + \dots,
	\eal
\eeq
which usually allows for a rather simple computation of this quantity. For holomorphic $(0,p)$-forms the relation \eqref{eq_chirealdefinition} generalizes to an alternating sum of certain Hodge numbers $\smash{h^{p,q}\ce\dim H^{p,q}_{\bar\partial}(X)}$, i.e.~
\beq
  \chi(X;\cO_X) = \sum_{p=0}^{\mclap{\dim_\IC X}} (-1)^p\,  h^{p,0}(X) = \int_X {\rm Td}(X),
\eeq
whose value minus 1 is called the arithmetic genus or holomorphic Euler characteristic. If bundle-valued forms are taken into account as well, we obtain the Hirzebruch-Riemann-Roch theorem
\beq
  \chi(X;V) = \sum_{q=0}^{\mclap{\dim_\IC X}} (-1)^q\,  h^{q}(X;V) = \int_X {\rm ch}(V)\, {\rm Td}(X).
\eeq
Note that for line bundles $L$ over $X$ the Chern character ${\rm ch}(L)$ simplifies to
\beq
  {\rm ch}(L) = \mathrm{e}^{c_1(L)} = \sum_{k=0}^\infty \frac{c_1(L)^k}{k!} = 1+c_1(L) + \frac{1}{2}c_1(L)^2 + \dots
\eeq

Some important special cases include the tangent bundle $T_X$, the cotangent bundle $T_X^*\cong\Omega^1_X$ and its antisymmetric tensor product $\Lambda^pT^*_X\cong\Omega^p_X$. Note that in general the Hirzebruch genera
\beq
  \chi_p \ce \sum_{q=0}^{\mclap{\dim_\IC X}} (-1)^q h^{p,q}(X) = \int_X {\rm ch}(\Omega^p) {\rm Td}(X)
\eeq
correspond to the alternating sum of the diagonal elements of the Hodge diamond, e.g.~for a K\"ahler 3-fold we have
\beq
	\begin{array}{ccccccccc}
	                  &         &                 & h^{0,0} &                  &          &                           &          & \\
		              	&         & \mclap{h^{1,0}} &         & \mclap{h^{0,1}}  &          &                           &          & \\
			              & h^{2,0} &                 & h^{1,1} &                  & h^{0,2}  &                           &          & \\
		\mclap{h^{3,0}} &         & \mclap{h^{2,1}} &         & \mclap{h^{1,2}}  &          & \mclap{h^{0,3}}           &          & \\
			              & h^{2,0} &                 & h^{1,1} &                  & h^{0,2}  &                           & \nwarrow & \\
			              &         & \mclap{h^{1,0}} &         & \mclap{h^{0,1}}  &          & \mclap{\nwarrow}          &          & \mclap{\chi_0} \\
	                  &         &                 & h^{0,0} &                  & \nwarrow &                           & \mclap{\chi_1} & \\
										&         &                 &         & \mclap{\nwarrow} &          & \mclap{-\chi_1} &          & \\
										&         &                 &         &                  & \mclap{-\chi_0} &               &          &
	\end{array}
\eeq
where $\chi_0 = \chi(X;\cO_X)$ and $\chi_1=\chi(X;\Omega^1_X)$.


\clearpage
%


\end{document}